\documentclass[aps,pra,twocolumn,superscriptaddress,showpacs]{revtex4}

\usepackage{amsmath,amssymb}
\usepackage{graphicx}
\usepackage[hypertex,breaklinks=true,colorlinks=false]{hyperref}
\def\rnum#1{\expandafter{%
\romannumeral #1}}
\def\Rnum#1{\uppercase\expandafter{%
\romannumeral #1}}

\newcommand{\PRL}[3]{Phys. Rev. Lett. {\bf #1},
\href{http://link.aps.org/abstract/PRL/v#1/e#2}{#2} (#3)}

\newcommand{\PRA}[3]{Phys. Rev. A {\bf #1},
\href{http://link.aps.org/abstract/PRA/v#1/e#2}{#2} (#3)}

\newcommand{\PRAR}[3]{Phys. Rev. A {\bf #1},
\href{http://link.aps.org/abstract/PRA/v#1/e#2}{#2(R)} (#3)}
\newcommand{\PRB}[3]{Phys. Rev. B {\bf #1},
\href{http://link.aps.org/abstract/PRB/v#1/e#2}{#2} (#3)}

\newcommand{\PRBR}[3]{Phys. Rev. B {\bf #1},
\href{http://link.aps.org/abstract/PRB/v#1/e#2}{#2(R)} (#3)}

\newcommand{\RMP}[3]{Rev. Mod. Phys. {\bf #1}, 
\href{http://link.aps.org/doi/10.1103/RevModPhys.#1.#2}{#2} (#3)}

\begin{document}

%\preprint{}

\title{Spontaneous population imbalance 
in two-component Bose and Fermi gases}


\author{Shintaro Takayoshi}
%\thanks{}
\affiliation{Institute for Solid State Physics, University of Tokyo, 
Kashiwa, Chiba 277-8581, Japan}
\author{Masahiro Sato}
\affiliation{Condensed Matter Theory Laboratory, RIKEN, Wako, Saitama 351-0198, Japan}
\author{Shunsuke Furukawa}
\affiliation{Department of Physics, University of Toronto, Toronto, Ontario M5S 1A7, Canada}
\affiliation{Condensed Matter Theory Laboratory, RIKEN, Wako, Saitama 351-0198, Japan}


\date{\today}

\begin{abstract}
We study two-component (or pseudo-spin-$\frac{1}{2}$) 
Bose or Fermi gases in one dimension, 
in which particles are convertible between the components. 
Through bosonization and numerical analyses of a simple lattice model, 
we demonstrate that, in such gases, a strong intercomponent repulsion induces 
spontaneous population imbalance between the components, 
namely, the ferromagnetism of the pseudo spins.  
The imbalanced phase contains gapless charge excitations characterized 
as a Tomonaga-Luttinger liquid and gapped spin excitations. 
% The imbalanced phase is characterized as a one-component 
% Tomonaga-Luttinger liquid, 
% with gapless and gapped excitations 
% in the charge and spin sectors, respectively. 
We uncover a crucial effect of the intercomponent particle hopping 
on the transition to the imbalanced phase. 
In the absence of this hopping, the transition is of first order. 
At the transition point, the energy spectrum reveals certain degeneracy indicative of an emergent $SU(2)$ symmetry. 
With an infinitesimal intercomponent hopping, the transition 
becomes of Ising type. 
We determine the phase diagram of the model accurately 
and test the reliability of the weak-coupling bosonization formalism. 
\end{abstract}

%\keywords{}

\maketitle

%\section{}
%\subsection{}
%\subsubsection{}
%\section{Introduction}

%%%%%%%%%%%%%%%%%%%%%%%%%%%%%%%%%%%%%%%%%%%%%%%%%
\section{Introduction}
%%%%%%%%%%%%%%%%%%%%%%%%%%%%%%%%%%%%%%%%%%%%%%%%%

%======================================
%- Cold atomic gas, creation of 1D, multicomponent

\begin{figure}
\begin{center}
\includegraphics[clip,scale=0.45]{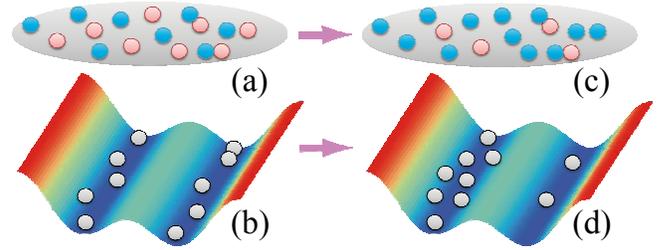}
\caption{(Color online) 
Illustrations of 1D two-component gases. 
The two components represent two internal states of atoms 
[shown by different colors in (a) and (c)]. 
Even without internal states,  
a double-channel trap potential can produce a similar situation 
[(b) and (d)]. 
A strong repulsion between the components 
induces the population imbalance [(c) and (d)].} 
\label{fig:ferro_illust}
\end{center}
\end{figure}

Ultracold atomic gases offer highly controllable laboratories 
for testing and exploring novel many-body phenomena in interacting 
systems \cite{Bloch08,Giorgini08}. 
One line of current interest is to confine atoms to highly elongated traps, 
effectively creating one-dimensional (1D) systems. 
In 1D interacting systems, the elementary excitations are collective modes, 
and the intuition based on the free-particle theory breaks down. 
For the spinless (one-component) case, theory predicts 
the equivalence of Bose and Fermi gases; 
both are described by the Tomonaga-Luttinger liquid (TLL) theory 
at low energies \cite{Giamarchi}. 
As a hallmark example, fermionization of bosons has been observed 
in a Bose gas of $^{87}$Rb atoms tuned into a strongly repulsive 
regime \cite{Kinoshita04,Paredes04}. 
Another frontier of activity is the creation of multicomponent gases 
using different internal states of atoms or using different species of atoms. 
If we limit our attention to 1D systems, 
a two-component Bose gas composed of the two hyperfine states of $^{87}$Rb 
has been confined in a 1D trap \cite{McGuirk02,Widera08}.
Even without internal states, one can load the atoms 
in a double-channel trap \cite{Hofferberth06,Hofferberth07,Jo07} [Fig.~\ref{fig:ferro_illust}(b)]
or on a ladder-type lattice \cite{Ripoll04,Strabley06}, 
effectively creating a 1D two-component gas.  
These two-component gases are expected to display 
a variety of phases depending on the intra- and intercomponent interactions. 
In a double-channel or ladder structure, in particular, 
the magnitudes of these two interactions would be 
different essentially and controllable separately. 

%##################
%##################

%======================================
%- Spontaneous population imbalance

% The multicomponent gases are expected to display 
% a variety of phases and phenomena, like a multiple of magnetic 
% orders induced by electron spins in solid. In this paper, 
% we study the two-component (i.e., pseudo spin-$\frac{1}{2}$) 
% Bose or Fermi gases in one dimension as the simplest many-body system 
% with internal dgrees of freedom. In particular, we focus on 
% a quantum phase transition caused by a strong inter-component repulsion. 

In this paper, we study the two-component Bose or Fermi gases in one dimension 
and analyze a quantum phase transition induced by a strong intercomponent repulsion. 
In two-component gases consisting of two species of atoms, 
it has been argued that a strong intercomponent repulsion 
induces the phase separation (the demixing) of the species
(see, e.g., Refs.~\onlinecite{Ho,Cazalilla_Ho,Pollet06,Mathey07,Mishra07,Hu_Clark,Kolezhuk}). 
Here we consider an analogous instability 
in the different situation where the two components 
represent two (internal) states of single-species atoms 
and thus particles are mutually {\it convertible} between the components. 
The system has a $\mathbb{Z}_2$ symmetry 
with respect to the interchange of the two components. 
In this case, it is expected that, under a strong intercomponent repulsion, 
a single component can dominate the whole system (population imbalance) 
as in Figs.~\ref{fig:ferro_illust}(c) and \ref{fig:ferro_illust}(d). 
The $\mathbb{Z}_2$ symmetry is spontaneously broken while the translational symmetry is retained. 
This ordering may also be viewed as ferromagnetism  
if we identify the two components with the pseudo spin $\frac{1}{2}$. 
A notable point in the present setting is 
that differently from the case of two-species mixtures, 
an intercomponent particle hopping can exist and violate 
the separate conservation of particle number in each component. 
It is also worth noting that the mechanism of spontaneous 
imbalance is in fact ubiquitous in nature. It underlies 
the ferromagnetism in $U(1)$-symmetric itinerant electrons \cite{Yang}, 
the vector chiral order in some frustrated 
magnets~\cite{Kolezhuk_Vekua,Sato_Sakai}, and 
the spontaneous rotation of a trapped Bose gas \cite{Tokuno_Sato}. 
This mechanism can occur both on a lattice and 
in a continuum since it does not involve any crystallization.

% In the demixing phase of a two-species gas, each separated region 
% can be described by the theory for a one-component gas, i.e., 
% the standard TLL theory. However, the $\mathbb{Z}_2$-broken phase 
% would not be captured by the TLL theory and deserves to be investigated. 
% In addition, several phenomena related to the above 
% spontaneous population imbalance are predicted to occur 
% in various fields of many-body physics: 
% We note that the spontaneous imbalance does not involve 
% any crystallization and can occur both on a lattice and in a continuum. 

%======================================
%- Theoretical understanding of spontaneous imbalance

In spite of their simplicity and ubiquity, 
the basic properties of the spontaneous imbalance and related phenomena 
have been rather poorly understood. 
Mean-field analyses of coupled Gross-Pitaevskii equations \cite{Ho} 
cannot capture the effects of strong quantum fluctuations in one dimension. 
A study beyond the mean-field argument has been done based 
on the weak-coupling bosonization formalism~\cite{Cazalilla_Ho}. 
In this formalism, each component is described as a TLL, 
and the intercomponent coupling is perturbatively treated. 
Then, as will be explained in Sec.~\ref{Sec3},   
the spontaneous imbalance is predicted to occur at a point 
where one of the TLL parameters diverges and this formalism breaks down \cite{Cazalilla_Ho}.
Therefore, this formalism cannot be used to describe the imbalanced phase 
nor the transition to it. 

Recently, 1D spin-polarized (ferromagnetic) Bose gases 
in a strongly repulsive regime 
have been studied actively using integrable models \cite{Lieb, Fuchs} 
and effective field theories \cite{Akhanjee_Tserkovnyak,Zvonarev,Matveev_Furusaki,Kamenev_Glazman}. 
In these studies, the original Hamiltonian has $SU(2)$ symmetry in terms of the (pseudo) spins, 
and this symmetry is spontaneously broken. 
In this setting, the excitations consist of a gapless charge mode characterized as a TLL and 
a gapless spin wave mode with a quadratic dispersion.  
In contrast, in our present setting, the (pseudo) spin $SU(2)$ symmetry 
is reduced to $\mathbb{Z}_2$. 
The low-energy excitation structure in such a reduced-symmetry case 
has not been addressed. 
Furthermore, these studies of $SU(2)$-symmetric systems focus only 
on the properties of the fully polarized state,  
and the nature of the ferromagnetic (population-imbalance) transition 
has not been discussed. 

%On the other hand, 
% Recently, 
% pseudo-spin-$\frac{1}{2}$ $SU(2)$-symmetric Bose gases 
% in a strongly repulsive regime 
% have been studied actively using integrable models \cite{Lieb, Fuchs} 
% and effective field theories \cite{Akhanjee_Tserkovnyak,Zvonarev,Matveev_Furusaki,Kamenev_Glazman}. 
% The ground state is known to be fully polarized and as a result, 
% the spin $SU(2)$ symmetry is spontaneously broken. 
% The charge mode obeys a TLL and spin excitations are described by 
% a gapless spin wave with a quadratic dispersion. 
% These $SU(2)$-symmetric models, however, are always in the 
% fully-polarized phase, and do not exhibit any phase transition. 
% Therefore, their results are not sufficient to understand 
% the imbalanced phase and the transition to it in our 
% $\mathbb{Z}_2$-symmetric setting. 
% We also note that when two-component stems from 
% certain internal degrees of freedom independent of true spin, 
% as in a double-channel trap of Fig.~\ref{fig:ferro_illust}(b) and (d), 
% the $\mathbb{Z}_2$-symmetric setting is more natural than 
% the $SU(2)$-symmetric models. 

%======================================
%- Our work

In this paper, we analyze a simple lattice model of a 1D two-component 
Bose or Fermi gas
to address the basic properties of the spontaneous population imbalance. 
We use two numerical methods, exact diagonalization and infinite 
time-evolving block decimation \cite{Vidal}, 
in efficient manners to go beyond the existing effective theories. 
We determine the phase diagram accurately and 
test the reliability of the bosonization prediction. 
Using the scaling of the entanglement entropy, 
we demonstrate that the low energy physics of the imbalanced phase is 
described by a one-component TLL, 
indicating that the excitations consist of a gapless charge mode and a gapped spin mode. 
We uncover a crucial effect of the intercomponent particle hopping 
on the nature of the transition. 
Perturbation theory from the strong-coupling limit in the half-filled case 
provides a qualitative understanding of the phase diagram and the nature of the transition. 

% For this purpose, we analyze a simple lattice model of a 1D two-component Bose or Fermi gas. 
%Note that such a hopping is absent in two-component gases consisting of different atoms. 

%%%%%%%%%%%%%%%%%%%%%%%%%%%%%%%%%%%%%%%%%%%%%%%%%
\section{Model}\label{Sec2}
%%%%%%%%%%%%%%%%%%%%%%%%%%%%%%%%%%%%%%%%%%%%%%%%%

%======================================
%- bosonic model

We start from a two-component Bose gas model 
on a 1D lattice defined by the Hamiltonian 
\begin{align}
{\cal H}=&\sum_{r=1,2}\sum_{j=1}^L [-t(b_{r,j}^\dag b_{r,j+1}+{\rm H.c.})
+Vn_{r,j}n_{r,j+1}\nonumber\\
         &~~~~~~~~~~~+U_{0} n_{r,j}(n_{r,j}-1)- \mu n_{r,j}]\nonumber\\
         &+\sum_{j=1}^L [-t'(b_{1,j}^\dag b_{2,j}+{\rm H.c.})+Un_{1,j}n_{2,j}],
\label{eq:Ham}
\end{align}
where $b_{r,j}$ is a bosonic annihilation operator 
at site $j$ in the $r$th component, 
and $n_{r,j}=b_{r,j}^\dagger b_{r,j}$ is the number operator defined from it. 
The first and second lines represent hopping and potential terms in each component, 
and the third represents those between the components. 
We set $t\ge 0$ and $t'\ge 0$. 
(The choices of the signs are arbitrary under gauge transformation.) 
For simplicity, we consider the hard-core limit $U_0 \to \infty$, 
where two particles in the {\it same} component cannot occupy 
the same site $j$ (but those in {\it different} components can). 

%======================================
%- Correspondence to the fermionic model

We are also interested in the two-component fermionic model  
which is defined by replacing all the bosonic operators 
$b_{r,j}$ in Eq.~\eqref{eq:Ham} 
by fermionic operators $f_{r,j}$. 
In the fermionic model, the hard-core interaction $U_0$ automatically 
drops out. When $t'=0$, the hard-core bosonic model is equivalent to the fermionic one 
via the Jordan-Wigner transformation:
\begin{subequations}
\begin{align}
 f_{1,j} &= \exp\left[ i\pi \sum_{l=1}^{j-1} n_{1,l} \right] b_{1,j},\\
 f_{2,j} &= \exp\left[ i\pi \left( \sum_{l=1}^L n_{1,l}+\sum_{l=1}^{j-1}n_{2,l} \right) \right] b_{2,j}, 
\end{align}
\end{subequations}
where the ``string'' part runs first in the first component and then in the second component. 
Because of this correspondence, 
the bosonic and fermionic models can be analyzed in parallel for $t'=0$. 
At the special point $t'=V=0$, 
the model~(\ref{eq:Ham}) is equivalent to the solvable fermionic Hubbard chain, 
where the labels $r=1$ and $2$ are identified with spin-up and spin-down states, respectively. 
In this case, the population imbalance is known not to occur. 

When $t'\ne 0$, the bosonic and fermionic models are no longer equivalent. 
For the following reasons, however, they are expected to display essentially the same physics. 
Specifically, we consider a different Jordan-Wigner transformation:
\begin{subequations}
\begin{align}
 f_{1,j} &= \exp\left[ i\pi \sum_{l=1}^{j-1} (n_{1,l}+n_{2,l}) \right] b_{1,j},\\
 f_{2,j} &= \exp\left[ i\pi \left( \sum_{l=1}^j n_{1,l}+\sum_{l=1}^{j-1}n_{2,l} \right) \right] b_{2,j},
\end{align}
\end{subequations}
where the ``string'' part now runs alternately between the two components. 
Under this transformation, the intracomponent hopping terms are transformed as
\begin{subequations}
\label{eq:bb-ff}
\begin{align}
 b_{1,j}^\dagger b_{1,j+1} &= e^{i\pi n_{2,j}} f_{1,j}^\dagger f_{1,j+1},
\label{eq:bb-ff1} \\
 b_{2,j}^\dagger b_{2,j+1} &= e^{i\pi n_{1,j+1}} f_{2,j}^\dagger f_{2,j+1}, 
\label{eq:bb-ff2}
\end{align}
\end{subequations}
while other terms in Eq.~\eqref{eq:Ham} retain the same form. 
In the low- (respectively high-) density limit, the phase factors $e^{i\pi n_{r,l}}$ in 
Eq.~\eqref{eq:bb-ff} are fixed essentially at unity (respectively $-1$).  
At half-filling and for strong intercomponent repulsion, 
the intracomponent hopping $t$ does not contribute 
in the first-order perturbation theory, 
as explained in Sec.~\ref{sec:perturbation}.  
Therefore, at least in these cases, 
the bosonic and fermionic models lead to the same physics.

%%%%%%%%%%%%%%%%%%%%%%%%%%%%%%%%%%%%%%%%%%%%%%%%%
\section{Weak-coupling theory}\label{Sec3}
%%%%%%%%%%%%%%%%%%%%%%%%%%%%%%%%%%%%%%%%%%%%%%%%%

We formulate a weak-coupling bosonization theory 
for the hard-core boson model \eqref{eq:Ham}, 
and discuss the instability of the two-component TLLs following Ref.~\onlinecite{Cazalilla_Ho}. 
When $t'=U=0$, the model decouples into two independent Bose gases, 
each equivalent to a solvable spin-$\frac{1}{2}$ XXZ chain 
in a magnetic field. 
For $-2<V$ and $0<\langle n_{r,j}\rangle<1$, 
each component $r(=1,2)$ obeys a TLL described by the Hamiltonian \cite{Giamarchi}
%the low-energy physics is described as 
%which is equal to a conformal field theory with central charge 
%$c=1$~\cite{Francesco}. 
%The TLL Hamiltonian is
\begin{equation}
{\cal H}^{\rm eff}_{r}=
\int{\rm d}x ~\frac{v}{2} \left[
 K(\partial_{x}\theta_{r})^{2}
+K^{-1}(\partial_{x}\phi_{r})^{2}
 \right],
\label{eq:Hamdecoup}
\end{equation}
where $\theta_{r}$ and $\phi_{r}$ are a dual pair of scalar fields, 
and $x=ja_0$ with $a_0$ being the lattice spacing. 
The group velocity $v$ and the TLL parameter $K$ can be determined from Bethe ansatz \cite{Giamarchi}. 
%by numerically solving integral equations \cite{Bogoliubov86}.
To treat $t'$ and $U$ terms as perturbations, 
we use the following bosonization formulas: 
\begin{subequations}
\begin{align}
n_{r,j} &= \rho_{0}-a_{0}\partial_{x}\phi_{r}(x)/\sqrt{\pi}+\cdots, \\
b_{r,j}^{\dagger} &= {\rm exp}[-{\rm i}\sqrt{\pi}\theta_{r}(x)](B_{0}+\cdots),
\end{align}
\end{subequations}
where $\rho_{0}=\langle n_{r,j}\rangle$ is the averaged density  
and $B_{0}$ is a nonuniversal constant. 
%which has been evaluated in Ref.~\onlinecite{Hikihara_Furusaki}.
Introducing new bosonic fields 
$\phi_{\pm}=(\phi_{1}\pm\phi_{2})/\sqrt{2}$ and 
$\theta_{\pm}=(\theta_{1}\pm\theta_{2})/\sqrt{2}$, 
we obtain the effective Hamiltonian for Eq.~(\ref{eq:Ham}), 
\begin{align}\label{eq:Heff}
{\cal H}^{\rm eff}=\int{\rm d}x\sum_{\alpha=\pm}\frac{v_{\alpha}}{2}
[K_{\alpha}(\partial_{x}\theta_{\alpha})^{2}
+K_{\alpha}^{-1}(\partial_{x}\phi_{\alpha})^{2}]& \nonumber
\\
-\frac{2}{a_{0}}B_{0}^{2}t'\cos(\sqrt{2\pi}\theta_{-})
-\sqrt{\frac{2}{\pi}}\rho_{0}U\partial_{x}\phi_{+}+\cdots,& %\nonumber
%&+(U\rho_{0}^{2}/a_{0})+\cdots,
\end{align}
with 
\begin{equation}\label{eq:vpm_Kpm}
v_{\pm}=v \left( 1\pm \frac{KUa_{0}}{\pi v} \right)^{\frac{1}{2}},~~ 
K_{\pm}=K\left(1\pm \frac{KUa_{0}}{\pi v}\right)^{-\frac{1}{2}}. 
\end{equation}
%The term $\partial_x \phi_+$ can be absorbed into the gaussian part 
%by shifting $\phi_+$ 
%and is neglected hereafter. 
Here we notice the effective separation of the two sectors,  
$(\phi_+,\theta_+)$ and $(\phi_-,\theta_-)$,  
which are respectively called the ``charge'' and ``spin'' sectors 
by analogy with the Hubbard chain. 
In the weak-coupling regime ($0\le U\ll v/a_0$), 
both the sectors remain gapless for $t'=0$. 
Finite $t'\neq 0$ opens a gap in the spin sector 
since  the vertex term 
$\cos(\sqrt{2\pi}\theta_{-})$ with scaling dimension $1/(2K_-)$ 
is always relevant. 
%In this region, the charge sector has no relevant perturbation 
%and is still described by a TLL. 
As $U$ increases, these estimates \eqref{eq:vpm_Kpm} 
indicate $v_{-}\to 0$ and $K_{-}\to\infty$ at 
\begin{equation}
U_{\rm c}=\pi v/(Ka_{0}).
\label{eq:Uc_bosonization}
\end{equation}
In other words, the coefficient $v_-K_-^{-1}$ of $(\partial_{x}\phi_-)^{2}$ 
in Eq.~\eqref{eq:Heff} changes sign at this point. 
This indicates the breakdown of the bosonization description 
in the spin sector. 
If we naively assume the existence of a term 
$(\partial_x \phi_-)^4$ with a positive coefficient 
in the effective Hamiltonian, a population-imbalanced state with 
$\Delta n=\langle n_{1,j}\rangle-\langle n_{2,j}\rangle
\approx -a_0 \sqrt{2/\pi} \langle \partial_x \phi_- \rangle  \neq 0$ 
is expected to appear along with the breakdown \cite{Yang,Tokuno_Sato}. 
Equation \eqref{eq:Uc_bosonization} gives a naive estimation of 
the transition point \cite{Cazalilla_Ho,comment1}. 
In the case where a two-component gas consists of two-species atoms,  
this breakdown corresponds to the demixing instability \cite{Cazalilla_Ho}.

%%%%%%%%%%%%%%%%%%%%%%%%%%%%%%%%%%%%%%%%%%%%%%%%%
\section{Numerical analyses}
%%%%%%%%%%%%%%%%%%%%%%%%%%%%%%%%%%%%%%%%%%%%%%%%%

To go beyond the weak-coupling theory and to address the 
strong-coupling regime, we employed two numerical methods for 
the bosonic model~(\ref{eq:Ham}): 
exact diagonalization (ED) and 
infinite time-evolving block decimation (iTEBD) \cite{Vidal}. 
The iTEBD method generates the ground state of an infinite system 
through the use of a matrix product state. 
The precision improves as we increase the matrix dimension $\chi$. 
To perform a calculation at a fixed filling, 
the chemical potential $\mu$ was iteratively tuned 
through a feedback control in each iTEBD step. 
To achieve better convergence of the order parameter, 
we first performed the iTEBD for large $U$, where the order parameter is large.  
%and then we adopted the obtained state as the initial state for slightly smaller $U$. 
Then we gradually decreased $U$ repeatedly using the obtained state 
as the initial state for the next $U$. 
We set $t=1$ as the energy unit hereafter. 

%%%%%%%%%%%%%%%%%%%%%%%%%%%%%%%%%%%%%%%%%%%%%%%%%
\subsection{$t'=0$ case}
%%%%%%%%%%%%%%%%%%%%%%%%%%%%%%%%%%%%%%%%%%%%%%%%%

\begin{figure}
\begin{center}
\includegraphics[clip,scale=0.85]{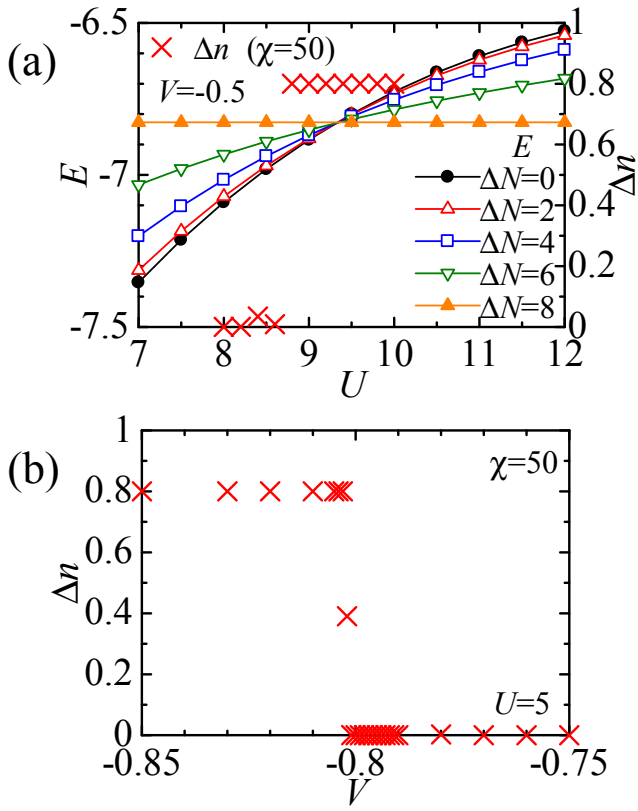}
\caption{(Color online) Numerical results for $t'=0$, $t=1$, and 
filling=0.4.
(a) The lowest energy levels for different $\Delta N$ 
obtained from ED and the density difference $\Delta n$ from iTEBD (with $\chi=50$), 
as functions of $U$. 
In ED calculations, the finite cluster of chain length $L=10$ was used. 
%The truncation matrix dimension of iTEBD is $\chi=50$. 
(b) $\Delta n$ obtained from iTEBD as a function of $V$. }
\label{fig:tp_0}
\end{center}
\end{figure}

Let us first analyze the case with zero intercomponent hopping $t'=0$, 
in which the bosonic model~(\ref{eq:Ham}) is equivalent to the fermionic one 
with the same Hamiltonian. 
In this case, the particle number of each component, $N_{r}=\sum_j n_{r,j}$, 
is a good quantum number. 
%is separately conserved. 
Therefore, for a given total particle number $N=N_1+N_2$, 
ED can be performed separately for each 
$\Delta N\equiv N_{1}-N_{2}=0,\pm 1,\dots,\pm N$. 
The lowest eigenenergy in each sector is plotted in Fig.~\ref{fig:tp_0}(a). 
We observe a direct change of the ground state 
from a uniform state $\Delta N=0$ 
to a fully imbalanced state $\Delta N=\pm N$ 
as we increase $U$. 
Such an abrupt change of $\Delta N$ indicates a first-order transition. 
%The change of the ground-state sector indicates that the transition is of first order.  
All the energy levels cross at the same point $U_c=9.34$, 
which suggests an emergent $SU(2)$ symmetry at the transition point. 
Such an emergent $SU(2)$ symmetry has also been predicted 
in renormalization group analysis \cite{Kolezhuk}. 
The recent strong-coupling theories for the $SU(2)$-symmetric case 
in Refs.~\onlinecite{Akhanjee_Tserkovnyak,Zvonarev,Matveev_Furusaki,Kamenev_Glazman} 
are expected to apply at this point. 
In Figs.~\ref{fig:tp_0}(a) and \ref{fig:tp_0}(b), the density difference $\Delta n$ 
evaluated by iTEBD shows a jump at a certain point, 
which also indicates a first-order transition. 
The transition points obtained from ED and iTEBD are slightly different, 
which could be mainly attributed to 
%a small finite-size effect in ED and to 
an inherent hysteresis in iTEBD around a first-order transition point. 
The level-crossing point in ED shows only a very small dependence on the system size 
except when the total density is close to zero or unity. 
Therefore, ED gives the better estimate of the transition point. 

%Since a finite size effect is small 
%except that the total density is near 0 or 1, 
%a level-crossing point of ED yield a better estimate for a transition point

%##################
%##################

%%%%%%%%%%%%%%%%%%%%%%%%%%%%%%%%%%%%%%%%%%%%%%%%%
\subsection{$t'\neq 0$ case}
%%%%%%%%%%%%%%%%%%%%%%%%%%%%%%%%%%%%%%%%%%%%%%%%%

%##################
\begin{figure}
\begin{center}
\includegraphics[clip,scale=0.85]{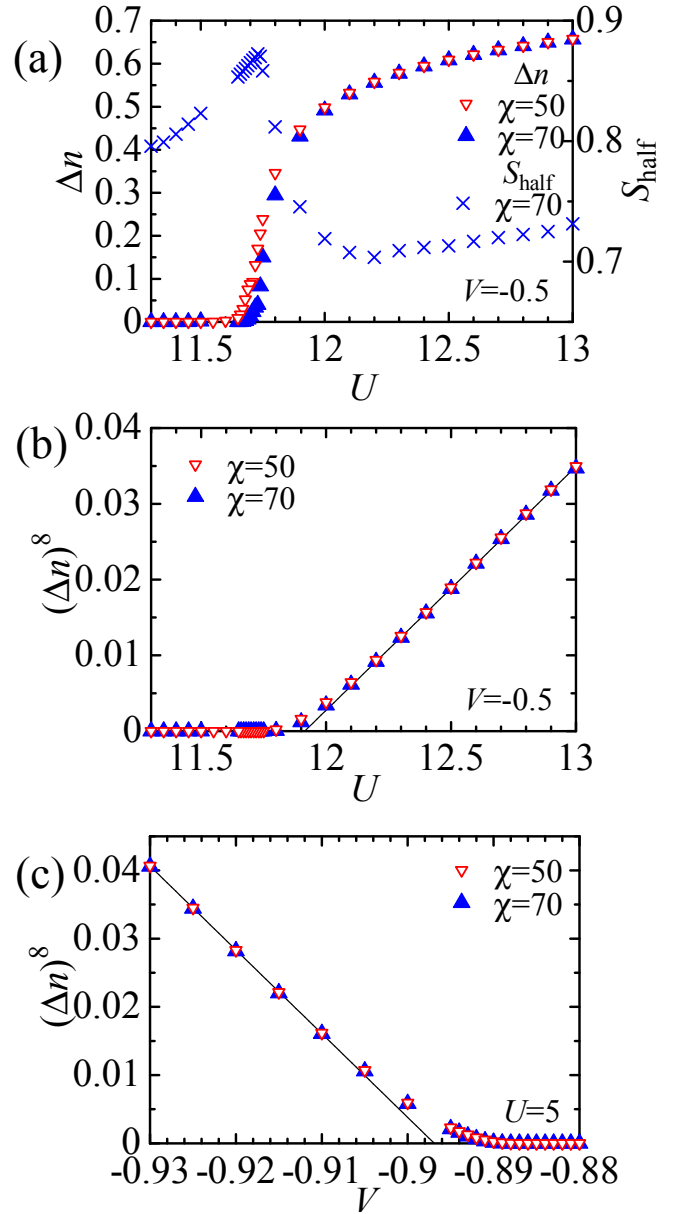}
\caption{(Color online) Numerical results for $t'=0.05$, $t=1$, and 
filling=0.4.
(a) Density difference $\Delta n$ and half-chain entanglement entropy 
$S_{\rm half}$
obtained from iTEBD, as a function of $U$. 
%with varying $U$. 
The onset of $\Delta n$ and the peak of $S_{\rm half}$ occur 
almost simultaneously. 
(b) $(\Delta n)^{8}$ versus $U$, and (c) $(\Delta n)^8$ versus $V$ 
near the transition point.
The black solid lines in (b) and (c) are 
the linear fits of the data 
%fitted from the numerical data 
in the imbalanced phase.} 
\label{fig:tp_non0}
\end{center}
\end{figure}
%##################

Now we analyze the effect of intercomponent hopping $t'\ne 0$, 
in which $N_{1,2}$ are no longer conserved separately. 
%The order parameter $\Delta n$ was calculated by iTEBD. 
%For $t'\neq 0$, the situation dramatically changes. 
As shown in Fig. \ref{fig:tp_non0}(a),
the order parameter
$\Delta n$ %\equiv\langle n_{1,j}\rangle-\langle n_{2,j}\rangle$
calculated with iTEBD 
grows continuously as a function of $U$, 
indicating a second-order transition. 
%i.e., the transition is a second-order type. 
The peak of half-chain entanglement entropy $S_{\rm half}$ 
[see Eq. (5) of Ref.~\onlinecite{Pollmann09} for its definition]
gives a reasonable estimate of the transition point. 
This quantity is known to diverge at a critical point \cite{Vidal03,Calabrese04}, 
although finite $\chi$ introduces a cutoff to the divergence \cite{Pollmann09}. 
In Figs.~\ref{fig:tp_non0}(b) and \ref{fig:tp_non0}(c), 
$(\Delta n)^{8}$ is plotted as a function of $U$ and $V$.
The data are well fitted by a linear function 
except in the very close vicinity of the transition point 
where $\chi$ dependence occurs. 
%almost independent of $\chi$ near the transition point. 
This result indicates the relation 
$\Delta n\propto (U-U_{\rm c})^{1/8}$ $[(V_{\rm c}-V)^{1/8}]$ 
along the $U$ $[V]$ axis, 
in agreement with the the critical exponent $\beta=1/8$ 
in the 2D Ising universality class~\cite{comment2}.

%%%%%%%%%%%%%%%%%%%%%%%%%%%%%%%%%%%%%%%%%%%%%%%%%
\subsection{Phase diagram}
%%%%%%%%%%%%%%%%%%%%%%%%%%%%%%%%%%%%%%%%%%%%%%%%%

\begin{figure}
\begin{center}
\includegraphics[clip,scale=0.85]{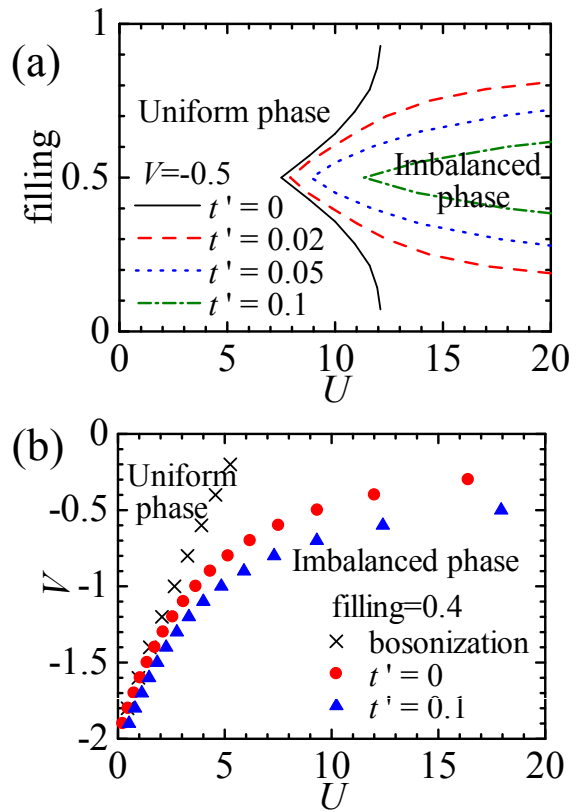}
\caption{(Color online) (a) %Global GS 
Phase diagram in the $U$-filling space for $V=-0.5$. 
We denote the phase boundaries by solid, dashed, dotted, and dash-dotted curves for 
$t'=0$, 0.02, 0.05, and 0.1, respectively. 
(b) Phase diagram in $U$-$V$ space for filling$=0.4$. 
%$V$ dependence of $U_{\rm c}$ derived from iTEBD. 
%The bosonization result corresponds to 
The Bosonization prediction (\ref{eq:Uc_bosonization}) of the phase boundary 
is also shown for comparison.} 
\label{fig:phase_diag}
\end{center}
\end{figure}

The ground-state phase diagram is summarized in Fig.~\ref{fig:phase_diag}.
%The phase boundaries are drawn by solid, dashed, dotted and 
%dash-dotted lines for $t'=0$, 0.02, 0.05 and 0.1, respectively.
The left and right sides of the phase boundary correspond to the uniform TLL and imbalanced phases, respectively. 
The transition points $U_{\rm c}$ were determined by 
using the level-crossing point for $t'=0$
and the peak of $S_{\rm half}$ for $t'\neq 0$.
%while the peak of half-chain entanglement entropy $S_{\rm half}$ is used for $t\neq 0$. 
In Fig.~\ref{fig:phase_diag}(a), the phase diagram is symmetric under $n\to 1-n$ 
because of the particle-hole symmetry in the hard-core model. 
It is found that $U_{\rm c}$ is shifted to larger values with increasing $t'$.
Namely, the intercomponent hopping diminishes the imbalanced state. 
Figure~\ref{fig:phase_diag}(b) shows the phase diagram in $U$-$V$ space. 
We observe $U_c\to 0$ as $V\to -2$. 
%When $V$ negatively increases to $-2$, 
%the bosonization result, Eq.~(\ref{eq:Uc_bosonization}), agrees with 
%the numerical one and $U_{\rm c}$ is close to zero. 
This is naturally expected
since, in the decoupled case ($t'=U=0$), 
the point $V/t=-2$ corresponds to the two ferromagnetic Heisenberg chains 
in the spin-system language with $S_{r,j}^z=n_{r,j}-1/2$. 
The bosonization prediction \eqref{eq:Uc_bosonization} 
and numerical data agree well in this limit. 
On the other hand, as $V$ is taken to zero, 
$U_c$ deviates from the bosonization prediction 
and tends to diverge. 
For $V\ge 0$, we did not observe a population imbalance, 
although the bosonization prediction \eqref{eq:Uc_bosonization} still indicates its occurrence. 
(We again note that the occurrence of the population imbalance 
can be disproved in the %$SU(2)$ 
integrable case $V=t'=0$.) 
This indicates intricate roles of $U$ and $V$ on the change of the TLL parameter $K_-$, 
which are not covered in the weak-coupling approach.

%##################
%##################

%%%%%%%%%%%%%%%%%%%%%%%%%%%%%%%%%%%%%%%%%%%%%%%%%
\section{Perturbation theory in the strong-coupling regime}
\label{sec:perturbation}
%%%%%%%%%%%%%%%%%%%%%%%%%%%%%%%%%%%%%%%%%%%%%%%%%

To gain a deeper understanding of the strong-coupling regime, 
we formulate a perturbation theory 
in the half-filled case $\langle n_{1,j}\rangle+\langle n_{2,j}\rangle=1$. 
In this special case, the charge sector is gapped out 
and we can thus single out a simple structure in the spin sector. 
(This procedure is analogous to the derivation of the Heisenberg model 
from the Hubbard model for a strong on-site repulsion.) 
In the limit $t/U,t'/U,V/U\to 0$, the system decouples into independent sites. 
%When $t=t'=V=0$, the system decouples into independent sites. 
Each site $j$ has doubly degenerate ground states, 
$|\!\Uparrow\rangle_{j}\equiv|n_{1,j}=1,n_{2,j}=0\rangle$ 
and $|\!\Downarrow\rangle_{j}\equiv|n_{1,j}=0,n_{2,j}=1\rangle$. 
%each of which contains a single particle occupying the state $r=1$ and $2$ respectively. 
We use these as the basis of the Hilbert space, 
and we treat $t$, $t'$, and $V$ terms as perturbations. 
%Since the ground states are degenerate, 
First-order perturbation theory is equal to acting the projection operator 
$P\equiv \prod_{j} (|\!\Uparrow\rangle_{j\: j}\langle\Uparrow\!|
+|\!\Downarrow\rangle_{j\: j}\langle\Downarrow\!|)$
on both sides of $t$, $t'$, and $V$ terms in Eq.~(\ref{eq:Ham}).
For both the bosonic and fermionic models, the same effective 
Hamiltonian is obtained as
\begin{equation}
{\cal H}_{\rm eff}^{(1)}= \sum_{j}(2VT_{j}^{z}T_{j+1}^{z}-2t'T_{j}^{x}),
\label{eq:Hami_stronglim}
\end{equation}
where $T_j^z$ and $T_j^x$ are pseudo-spin-$\frac12$ operators defined as
\begin{equation}
\begin{split}
T_{j}^{z}\equiv&\frac{1}{2}(
|\!\Uparrow\rangle_{j\: j}\langle\Uparrow\!|
-|\!\Downarrow\rangle_{j\: j}\langle\Downarrow\!|)\\
T_{j}^{x}\equiv&\frac{1}{2}(
|\!\Downarrow\rangle_{j\: j}\langle\Uparrow\!|+
|\!\Uparrow\rangle_{j\: j}\langle\Downarrow\!|).
\end{split}
\end{equation}
%satisfying 
%$T_j^z |\!\Uparrow\rangle_{j} = \frac12 |\!\Uparrow\rangle_{j}$, 
%$T_j^z |\!\Downarrow\rangle_{j} = -\frac12 |\!\Downarrow\rangle_{j}$, and
%$T_j^x |\!\Uparrow\rangle_{j} = \frac12 |\!\Downarrow\rangle_{j}$. 
%defined in the basis of 
%$|\!\Uparrow\rangle_{j}$ and $|\!\Downarrow\rangle_{j}$. 

For $t'=0$, Eq.~\eqref{eq:Hami_stronglim} is a classical Ising model. 
A first-order transition at $V=0$ separates 
ferromagnetic ($V<0$) and antiferromagnetic ($V>0$) phases. 
The former is nothing but the population-imbalanced phase. 
For $t'> 0$, Eq.~\eqref{eq:Hami_stronglim} is equal to an Ising model in a transverse field, 
which is still solvable. 
Second-order transitions of Ising type separate 
ferromagnetic ($V<-2t'$), disordered ($-2t'<V<2t'$), and antiferromagnetic ($2t'<V$) phases. 
The presence of $t'$ abruptly changes the nature of the transition, 
consistent with the numerical results in Figs.~\ref{fig:tp_0} and \ref{fig:tp_non0}. 
Furthermore, the ferromagnetic phase diminishes as we increase $t'$, 
in accordance with Fig.~\ref{fig:phase_diag}. 
The appearance of the antiferromagnetic phase is due to a lattice effect specific to the half-filled case, 
and here we do not discuss it further. 
% Since the effective Hamiltonian (\ref{eq:Hami_stronglim}) is the same
% for both bosons and fermions,
% the properties of the ferromagnetic transitions observed here 
% is independent of the statistics of the particles.

%##################
\begin{figure}
\begin{center}
\includegraphics[clip,scale=0.85]{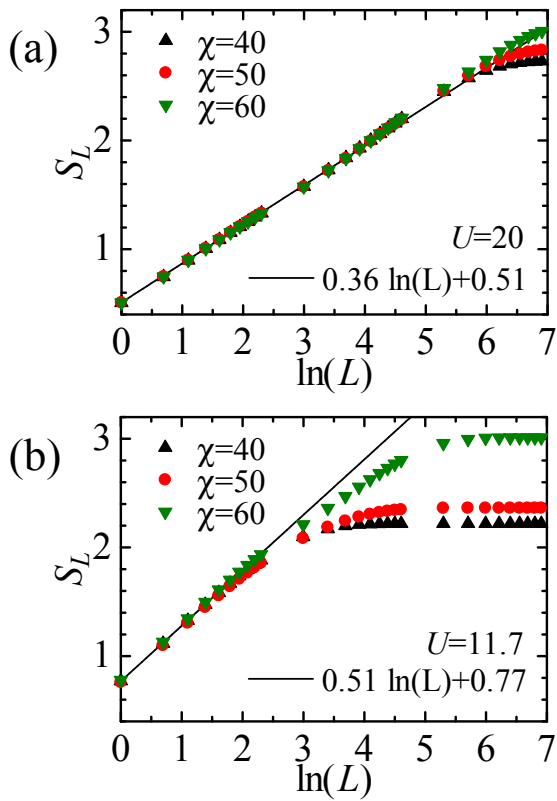}
\caption{(Color online) 
Finite-interval entanglement entropy $S_L$ calculated by iTEBD. 
%Entanglement entropy of the subregion with the chain length $L$. 
We set $t'=0.05$, $V=-0.5$, and filling=0.4.
(a) In the imbalanced phase, we obtain $c\approx 1$.
(b) Approximately at the transition point, we obtain $c\approx 1+1/2$.} 
\label{fig:entropy}
\end{center}
\end{figure}
%##################

%The Ising model shows that a negative large $V<0$ is needed for 
%the imbalanced state  

%because $|\Uparrow\;\rangle_{j}$ and $|\Downarrow\;\rangle_{j}$ represents
%$|0,1\rangle$ and $|1,0\rangle$, respectively. 

%%%%%%%%%%%%%%%%%%%%%%%%%%%%%%%%%%%%%%%%%%%%%%%%%
\section{Low-energy properties of the imbalanced phase}
%%%%%%%%%%%%%%%%%%%%%%%%%%%%%%%%%%%%%%%%%%%%%%%%%

We analyze the low-energy properties of the imbalanced phase. 
In the case of $t'=0$, it is clear that the low-energy physics is governed 
by a TLL in the charge sector since the fully polarized state is realized 
(see Fig.~\ref{fig:tp_0}). On the other hand, the case of $t'\neq 0$ deserves to 
be investigated. In the latter case, it is natural to consider the 
following two possibilities: 
(i) The effective separation of the charge and spin sectors still holds;  
the former is described by a TLL while the latter is gapped. 
(ii) The first chain ($r=1$) with a dense particle density and 
the second one ($r=2$) with a thin density separately form TLLs. 
%To distinguish between the two possibilities, 
To see which possibility is correct, 
we examine the scaling of the entanglement entropy $S_{L}$ of the ground state $|\Psi\rangle$. 
For a finite interval $\Omega$ of length $L$, it is defined as 
$S_{L}=-{\rm Tr}\rho_{\Omega}\ln\rho_{\Omega}$, 
where
$\rho_{\Omega}={\rm Tr}_{\bar{\Omega}}|\Psi\rangle\langle\Psi|$ 
is the reduced density matrix on $\Omega$ obtained by tracing out the exterior $\bar{\Omega}$.
%of the ground state $|\Psi\rangle$ 
% $|\Psi\rangle$ is the ground state of the system. 
%, and ${\bar{\Omega}}$ is the outside of $\Omega$. 
In 1D critical systems, this quantity enables one 
to determine the central charge $c$  
(an indicator of the number of gapless modes) 
through the formula of a universal scaling \cite{Vidal03,Calabrese04},  
%$S_{L}\approx\frac{c}{3}\log L+{\rm const}$. 
\begin{equation}
S_{L}\approx\frac{c}{3}\ln L+{\rm const}.
\label{eq:ent_Ldep}
\end{equation}
Figure~\ref{fig:entropy} shows $S_{L}$ calculated by combining 
iTEBD with the transfer matrix technique. 
Here, an initial state with a fixed number of particles was adopted 
to achieve better convergence of the wave function. 
In the imbalanced phase [Fig.~\ref{fig:entropy}(a)], 
$c\approx 1$ is obtained by fitting the numerical data with 
the scaling formula \eqref{eq:ent_Ldep}. 
This supports possibility (i). 
%It indicates that a one-component TLL survives and the spin sector is gapped.
%In terms of boson fields, $\phi_{-}$ sector has an energy gap
%while $\phi_{+}$ sector remains a gapless TLL with increasing $U$.
%Namely, a gapless liquid and the ferromagnetic order coexist in this phase. 
Namely, the dense and thin components cooperatively form a single TLL. 
This is strikingly different from a population-imbalanced state
%with fixed $N_1$ and $N_2\neq N_1$. 
extrinsically derived by simply applying a magnetic field 
(chemical potential difference) for pseudo spins \cite{comment3}.  
Figure~\ref{fig:entropy}(b) shows the result approximately at the second-order transition point $U_c$ for $t'\neq 0$. 
Although the data show some $\chi$ dependence for large $L$, 
the fitting using almost $\chi$-independent data for small $L$ yields $c\approx 1+1/2$. 
This result indicates that an Ising-type transition with $c=1/2$ 
occurs in the spin sector 
while the gapless mode with $c=1$ in the charge sector remains intact. 

% The entropy $S_L$ for $t'\neq 0$ is also calculated 
% near the transition point as shown in Fig. \ref{fig:entropy}(b).
% Although the convergence of iTEBD calculations
% is not so good near $U_{\rm c}$,
% the fitting line clearly suggests $c\approx 1+1/2$.
% This again supports that the transition belongs to
% the Ising class with $c=1/2$.
% Fixed initial states are used in iTEBD calculations of $S_{L}$
% for better convergence while random ones in other calculations.

%%%%%%%%%%%%%%%%%%%%%%%%%%%%%%%%%%%%%%%%%%%%%%%%%
\section{Conclusions}
%%%%%%%%%%%%%%%%%%%%%%%%%%%%%%%%%%%%%%%%%%%%%%%%%

We have studied a simple lattice model \eqref{eq:Ham} 
of a two-component quantum gas in the strong-coupling regime, 
using numerical and analytical methods. 
We demonstrated that the spontaneous population imbalance (ferromagnetism) occurs 
as we increase the intercomponent repulsion $U$. 
We completed the accurate phase diagrams in Fig.~\ref{fig:phase_diag}. 
It was found that the spontaneous imbalance occurs only for $V<0$ 
and that the intercomponent hopping $t'$ diminishes the imbalanced phase. 
While these intricate roles of $V$ and $t'$ cannot be covered in weak-coupling bosonization theory, 
perturbation theory from the strong-coupling limit for the half-filled case 
provides simple qualitative explanations of them. 
%The determined phase diagram is consistent with perturbation in the strong coupling region.

We have also uncovered the basic properties of the imbalanced phase and 
the transition to it. 
Using the scaling of the entanglement entropy, we demonstrated that 
the low-energy property in 
the imbalanced phase is governed by a one-component TLL, 
indicating the separation of gapless charge and gapped spin modes. 
This is in sharp contrast to the integrable fermionic Hubbard chain, 
where both the charge and spin sectors behave as TLLs 
even in the strong-coupling regime~\cite{Ogata90}. 
The transition to the imbalanced phase is 
of first order when $t'=0$ 
and of Ising type when $t'\neq 0$. 
In spite of this drastic effect in the spin sector, 
the $t'$ term does not spoil the gapless property of the charge sector. 
At the first-order transition point for $t'=0$, 
the energy spectrum reveals a certain degeneracy 
indicative of an emergent $SU(2)$ symmetry. 

Although our analyses are done mainly for the bosonic model, 
we expect that the fermionic model also displays essentially the same physics 
as we discussed in Sec~\ref{Sec2}. 
For $t'=0$, the bosonic and fermionic models are exactly equivalent. 
For $t'\ne 0$, the $t'$ term in the fermionic case has a different bosonization expression 
and thus a different scaling dimension from the bosonic case in Eq.~\eqref{eq:Heff}. 
We expect that, when this term is irrelevant (respectively relevant), 
the transition between the uniform TLL and imbalanced phases is of first order (respectively of Ising type). 
In particular, even in the presence of the $t'$ term, 
the bosonic and fermionic cases become asymptotically equivalent 
(i) in the low- and high-density limits 
and (ii) in the strong-coupling limit in the half-filled case. 

% Particularly, in the strongly inter-component repulsive regime, 
% the low-energy physics of the bosonic model~\ref{eq:Ham} 
% would be equal to that of the fermionic one 
% near the half-filled case and in the low- and high-density limits 
% (see Secs.~\ref{Sec2} and \ref{sec:perturbation}). 

Releasing the hard-core constraint 
and discussing the occurrence of the population imbalance 
in more realistic situations are interesting future directions. 
We expect that the basic features of the imbalanced phase and the transition  
uncovered in the present work are robust, 
irrespective of the microscopic details.

%======================================
%- Acknowledgement
S.T. and M.S. were supported by Grants-in-Aid by JSPS 
(Grant No.\ 09J08714) and for Scientific Research from MEXT 
(Grant No.\ 21740295 and No.\ 22014016), respectively.


\begin{thebibliography}{}
\bibitem{comment1}
Recently, an improved estimation of the transition point 
beyond simple bosonization was argued in Ref.~\onlinecite{Kolezhuk}.
\bibitem{comment2}
In Ref.~\onlinecite{Kolezhuk_Vekua}, an Ising-type 
transition was predicted to occur between the vector chiral phase 
and the nematic liquid phase 
(called the even-odd phase in Ref.~\onlinecite{Kolezhuk_Vekua}) 
in the zigzag spin-$\frac{1}{2}$ chain in a magnetic field. 
We comment that this Ising transition is essentially different 
from the one we found in the present case. 
In the bosonization formalism, different bosonic fields are locked 
in the nematic liquid and uniform TLL phases.
\bibitem{comment3}
Population imbalanced Fermi gases obtained by applying a magnetic field 
are currently an active topic and are reviewed in, e.g., Sec. IX of Ref.~\onlinecite{Giorgini08}.
\end{thebibliography}
\end{document}